# Facial Gesture Interfaces for Expression and Communication[*]

**Michael J. Lyons**
ATR IRC/MIS Labs
2-2-2 Hikaridai, Keihanna Science City, Kyoto 619-0288
mlyons@atr.jp

**Abstract -** Considerable effort has been devoted to the automatic extraction of information about action of the face from image sequences. Within the context of human-computer interaction (HCI) we may distinguish systems which allow expression from those which aim at recognition. Most of the work in facial action processing has been directed at automatically recognizing affect from facial actions. By contrast facial gesture interfaces, which respond to deliberate facial actions, have received comparatively little attention. This paper reviews several projects on vision-based interfaces which rely on facial action for intentional HCI. Applications to several domains are introduced, including text entry, artistic and musical expression and assistive technology for motor impaired users.

## 1   Introduction

Muscular action of the face plays an important role in much of human behaviour including speech, facial expression and gesture. Penfield [15] discovered that neural circuits associated with the hands and face, particularly the mouth, occupy a disproportionate share of sensory and motor cortical area. It is therefore not unreasonable to think that actions of face could play an important complementary or supplementary role to that played by the hands in machine-interaction.

While there is a considerable body of prior research on automatic facial expression recognition and lip reading, there has been relatively little work examining the possible role of the face in direct, intentional interactions with computers or other machines. This may be partly due to technological limitations: how can information about motor actions of the mouth be acquired in an non-encumbering, non-invasive fashion? With the extensive work on facial expression recognition over the past decade [13], however, vision-based methods now offer a realistic solution to this obstacle.

Recently we have been using vision-based methods to capture movement of the head and facial features and use these for intentional, direct interaction with computers. Two of the projects that will be reviewed below allow text entry involving motion of the head and/or mouth. A further two projects discussed in this work explore the concept of using motion of the mouth for artistic and musical expression. While our primary intention is to suggest that facial actions could provide an input channel for HCI which is parallel to and independent of action of the hands, it could also be of use for motor-impaired computer users.

The unusual nature of the idea of using the face for intentional interaction may be another factor in the relative dearth of precedent studies, however novelty or lack of familiarity of a concept should not deter research. In this paper we review several of our projects in this area to support the thesis that facial gesture HCI can be natural, useful, and fun.

## 2   Methods

We have chosen to concentrate on the most naturally expressive area of the face, the mouth, and apply simple, rapid, and robust vision methods to implement and test actual interactive systems. We have employed both face-tracking methods (Figure 1) and a headworn miniature camera (Figure 2) in the systems developed. Each has advantages and disadvantages. The face-tracking systems are non-invasive – the user does not need to wear any equipment or tracking labels. However they must face a camera. This is fine for desktop computing applications but becomes problematic for situations where mobility is important. With a headworn camera pointed at the face, the user is free to roam, without changing the image captured by the camera, modulo lighting changes.

Two distinct methods for face tracking have been used (Figure 1). We omit all but immediately salient details of the tracking methods and refer interested readers to source papers [4,5]. Both methods assume situations in which the user faces the camera approximately frontally (± ~30º). With one tracking method, abbreviated hereafter as *NP* (Nose Pointer) [4], the camera is slightly higher than the eye level of a (usually) seated user. With the other, *NF* (Nostril Finder) [5], the camera is positioned lower than the user's face, either resting on a desktop, or attached to a hand held device pointed upwards at the face.

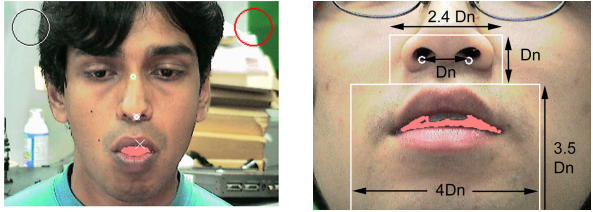

Figure 1. Face tracking systems used in our work: the nose pointer (*NP*) and nostril finder (*NF*).

With the *NP system*, tracking is initiated by blink detection followed by tracking of scale-insensitive structure in the vicinity of the eyes and estimation of eye positions. The nose tip is located as the brightest point in a region of interest determined by the eyes [6]. The eyes and nose determine a region of interest for segmentation of the shadow area in the open mouth (MSROI). When tracking is lost, it can be reinitiated with no manual intervention simply by blinking the eyes while looking at the computer screen. With *NP*, head movements can be used to control a cursor, while mouth movements can be used for discrete or continuous control of one or two parameters, for example entering mouth clicks [4].

With the *NF system*, initiation of tracking is semi-manual. The nostrils are positioned in the center top half of the image and a button pressed. The centers of the nostrils are located using a modified version of a algorithm introduced by Petajan [14]. The current nostril positions determine the search region at subsequent times, until tracking is re-initialized. The nostril positions are used to estimate the position, scale, and orientation of the MSROI. With *NF*, we use only the mouth controller for parameter output.

With the head-worn tracker the vision problem is simplified: with the camera positioned a few centimeters from the face, part or whole of the range of view is taken as the MSROI.

## 2.1 Mouth Controller

Under a wide range of lighting conditions an open mouth produces a shadow, which can be quite robustly segmented with thresholding on the intensity and red component of the pixels.

Region growing and connected components analysis, with selection of the largest blob are used to reduce noise. Dimensions of the bounding box or shape statistics, such as area, aspect ratio, or vertical and horizontal second moments are taken parameters to be controlled by varying mouth shape. These parameters are invariant to translations of the camera so that the inevitable motions of the headworn camera are do not strongly affect the signal. Principle components analysis of the shape of the blob can also be used to gain rotational invariance [10].

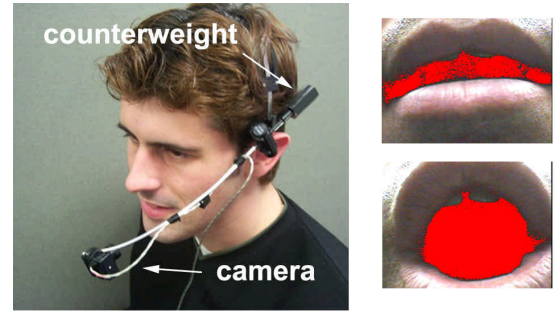

Figure 2. Head-worn camera and views of the mouth.

## 3 Control Tasks and User Studies

In studying interactive technologies it is important to not only assess the functioning of the technology but to study the human aspect of the interaction as well. For example, some facial actions are more natural and easily performed than others. It is simple enough to open and close the mouth by movements of the jaw and or lips or to stretch the corners of the mouth in a grin or grimace. Some facial actions, such as moving only one eyebrow, wriggling the nose or ears, are very difficult for most people.

To assess the degree of control afforded by vision-based capture of mouth action we have devised several novel usability tasks (Figure 3). The general principle is to map mouth shape features to shape parameters of geometric figures on a computer monitor. The user then moves the mouth to adjust the shape of the controlled figure until it matches a target shape.

With the *circle control task* [2] the area of the open mouth is used to control the size of a circle. The single parameter in this mapping is the gain of the mapping. We studied the effect of gain on the accuracy and precision of control of the size of the circle. It was found that lower gains result in better accuracy and precision of control for novice users, but that experts can perform with approximately the same accuracy for various gain settings. Near single pixel accuracy in the radius of the circle was achieved for a wide range of sizes of target circles. This corresponds to a signal/noise ratio of 60-70 dB under most conditions studied. Similar results were obtained with a single parameter (such as the mouth height or width) was used. The high signal to noise ratio reflects the that the controlled parameters depend on the statistics of a large number of pixels as well as the fact that the simple thresholding methods used are translation and rotation invariant.

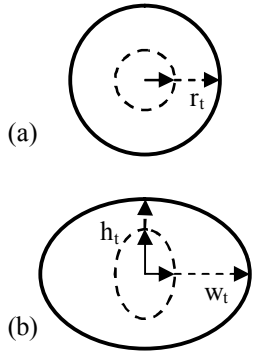

Figure 3. (a) circle and (b) ellipse control tasks

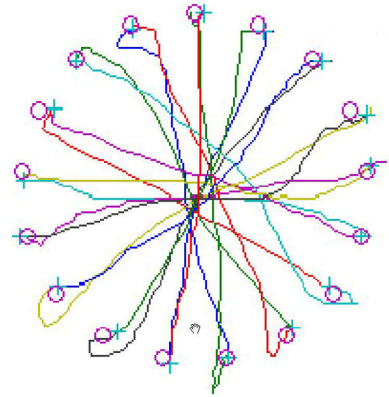

Figure 4. Tracks observed with the ISO multidirectional tapping task

With the *ellipse control task* the user adjusts the aspect ratio and open area of the mouth simultaneously, or, equivalently, the width and height of the open mouth. These control the width and height, respectively, of an ellipse drawn on the computer monitor. The user is required to match a target ellipse and hold the configuration for 3 seconds. This is the most difficult of the control tasks, and some configurations, for example a large value of height with a small value of width were observed to be difficult to control for some users.

With the *NP* system the user can position the cursor by pointing with the nose [4]. The gain of the system can be adjusted so that relatively small head movements are needed. We have studied the efficiency and usability of NP as an input method for cursor control using the ISO multidirectional tapping task [7]. This consists of a series of target acquisition tasks arranged around the circumference of a circle (Figure 4). The figure shows sample tracks input with the *NP* system during the experiment. Target acquisition times are used to calculate an information throughput efficiency of 2.0 bits/sec which is lower than the 5 bits/sec we measured using a mouse, but similar to values previously reported for joystick and trackball devices. These results were in agreement with earlier head movement studies using more intrusive non vision-based head pointing systems.

## 4 Metaphor, Mapping, Applications

It is less straightforward to assess which applications, and what mapping of action to effect within a given application will be attractive and conducive to creative expression for users. *Metaphor* has been found to be a useful concept in guiding the design of interface technology. A metaphor is a way of representing one experience or concept in terms of another [8]. Metaphor can help us to make new experiences or behaviours accessible by relating them to something we are already familiar with. The idea of positioning a cursor by pointing the nose in different directions is an example of a metaphor that makes *NP* easy to use. Another is the metaphor of controlling the size of a shape by varying the open area of the mouth.

A related question is: what kind of action-effect mappings result in comfortable, easy to learn interfaces? Should the output be proportional or inversely proportional to the measured control parameter? Should the proportionality be linear, nonlinear or even discrete or non-monotonic? Such issues are usual in the study of input devices [1]. A full discussion is beyond the scope of the current paper.

In the next sections I review several projects which used video-based acquisition of facial action for machine interaction purposes.

## 5 Mouth-Driven Musical Interface

Our first real-time interactive facial action system was a mouth-driven musical effects controller (Figure 5) [11]. The inspiration for this system came from observing the facial expressions made by some musicians while playing. It is also motivated by the role of the mouth in sound production in speech, singing, whistling, and humming. In early experiments we mapped movements of the eyebrows and mouth to control parameters of digital sound synthesis, but soon concentrated on the mouth as the most salient facial part for musical control. The first prototype included a vision-based face-tracking system, but we switched to a head-worn system to allow the musicians more mobility. The search for mappings of facial action to musical effect was guided by two types of metaphors. One is the relation of facial action to emotion. For example, stretching the corners of the mouth wide in a grimace is related to expressions of effort or pain. It is effective to map these to audio distortion, for example, a quality valued in many forms of folk and popular music. Sound production by the vocal tract also guided our search for mappings. For example, the effect known by the onomatopoeic term 'wah-wah' is essentially the audio effect obtained by voicing 'ah' while opening and closely the mouth. Vowel formant filters are also used widely in electronic music and are very naturally controlled by mouth shape. Exploration of these and other mappings in trials with several musicians with led us to conclude that action of the mouth is a very natural and compelling way to control musical effects. The mouth controller, or mouthesizer, as it is known, has been used in a public performance of electronic music [10].

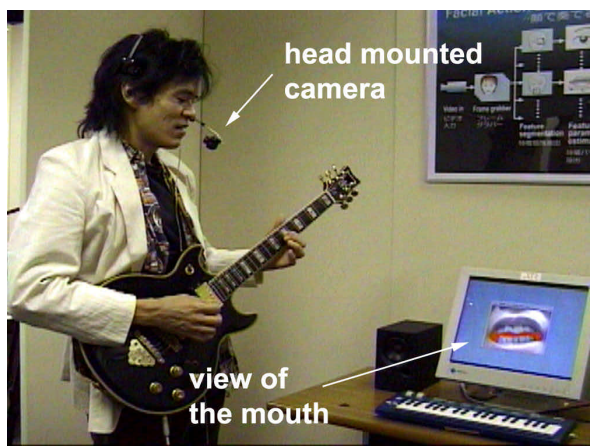

Figure 5. The Mouthesizer musical controller

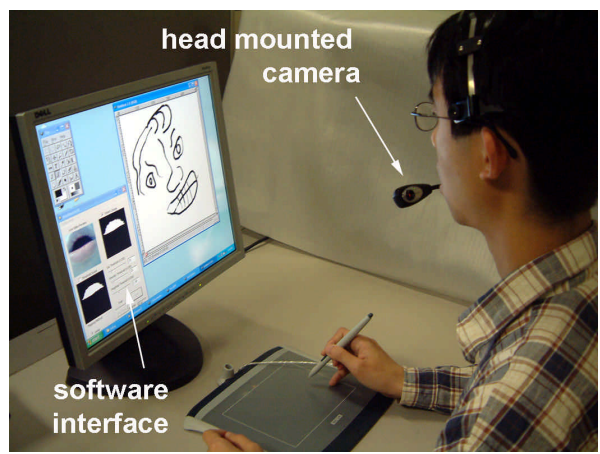

Figure 6. Drawing with Mouthbrush.

## 6 Digital Painting by Hand and Mouth

Some of the earliest paintings are thought to have been created by blowing pigment through a tube pointed at a surface such as a cave wall. Blowing paint through a tube was used in painting and decoration technique until the development of air brushes with air compressors. We have re-introduced action of the mouth into the creation of visual art in our Mouthbrush project [2]. With the Mouthbrush (Figure 6), the area of the open mouth is used to control qualities of the brush or paint in a digital drawing or painting program. The Open Source drawing program, Gimp, was modified to accept a parameter sent via TCP/IP and map it to brush properties such as size and hardness, and/or paint properties such as colour and opacity. This allows the artist facile and continuous control of brush or paint properties while the stylus is moved. Such properties are sometimes linked to stylus pressure with a tablet input device. The advantage of the mouth controller is that it allows an independent motor system to act as the hand moves. A number of amateur artists have tried our system, and it was found to be simple and engrossing way to create digital art. The mouthbrush was first implemented with the headworn camera system. More recently it has been implemented to work with a desktop camera and the *NF system*.

## 7 Text Entry

Writing is arguably the most fundamental human communication technology. Various human methods have been employed for the creation of written texts: for example hand writing with a stylus on clay or paper, various printing techniques, and electronic text entry using a QWERTY keyboard or 12 key telephone keypad. Recent growth in the popularity of mobile and handheld computing and led to increased exploration of novel and sometimes unusual methods for text entry. Most text entry methods involve motor action of the hands, though head movements and eye movements have been employed, primarily for text entry by disabled users [4,17].

### 7.1 Text Entry by Head Movements

Our group has explored use of a vision-based face-tracking system (the NP system) for non-encumbered text entry by head movements [4]. We used the Open Source Dasher system [17] which allows text entry by 2-D cursor movements. Combined with the NP system, relatively small head movements can be used to input text. Text entry was initiated and concluded in a hands-free fashion by opening and closing the mouth (corresponding to a mouth click). The measured rate of text entry was 7-12 words/minute depending on user expertise. This system could be useful for motor-impaired computer users.

### 7.2 Text Entry by Hand and Mouth

Some believe that speech recognition, perhaps coupled with automatic lip-reading will be the ultimate text entry method in the future. However, we are far from having a general speech-to-text interface. Robust, real-time lip-reading is also a distant goal. Moreover, fundamental human factors considerations indicate that speech recognition systems may not be desirable in many text creation situations [16]. It is, however, interesting to consider a more technically modest system whereby silent, intentional movements of the mouth are coupled with action of the hands for text entry. Since the mouth is involved in speech we may be able to make use of existing human expertise, reducing the effort required to learn or use a given text entry system. We have implemented two

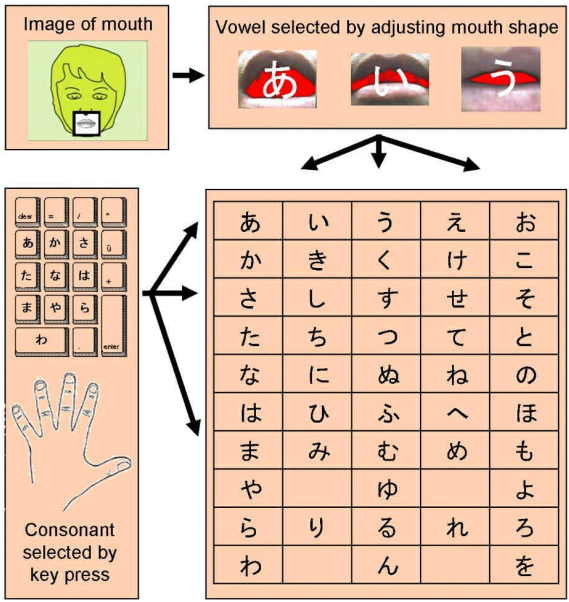

Figure 7. Hiragana input by hand and mouth

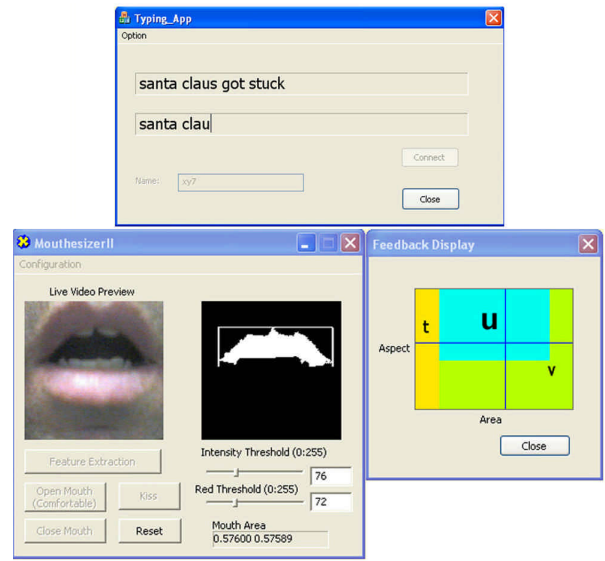

Figure 8. English text entry with MouthType.

such systems, one for Japanese text input and one for use with the Roman Alphabet [9].

Figure 7 shows a schematic of the MouthType system as used for Japanese text entry. To understand this system it is first necessary to introduce a few basics facts about the Japanese writing system [3]. Written Japanese is a mixture of two isomorphic syllabic writing systems (Hiragana and Katakana) and a logographic writing system (the kanji or Chinese characters). Most contemporary Japanese enter Hiragana on QWERTY keyboards by typing 2 letter roman spellings of the syllable. Conversion of syllabic elements to Hiragana takes places via menu systems, often using a predictive language model. To enter the 'su' syllable in the word 'sushi', for example, one types a 's' followed by an 'u'. In Japanese most syllables consist of a consonant followed by one of the five Japanese vowels (written 'a', 'i', 'u', 'e', 'o').

With mobile phone keypads the text input method actually corresponds more closely to the Japanese writing system that the text entry using a QWERTY keyboard. Consonants are mapped to individual keys, with one of five vowels being selected by multiple key presses of the same key (known as the 'multi-tap' text entry method). With the MouthType system, we preserve the mapping of consonants to the keys of the keypad but select the vowel using the mouth shape as acquired by a video camera pointed at the face. So to select the syllable 'ka', the user presses the key corresponding to the 'k' series of Hiragana, while shaping their mouth so as to silently pronounce the vowel 'a'. With this method all of the Hiragana can be selected by a single key press. We have implemented a complete system for Hiragana text entry, with provision for compound syllables and diacritic marks [9]. We compared the speed of text entry using this system with that measured for the usual multi-tap method of text entry and found that it is significantly faster. It was also found to be easy to understand and use. Little investment of time is needed to learn the system as it relies on pre-existing user expertise in shaping the mouth to pronounce the five vowels.

Similarly, mouth shape can be used to select among the multiple letters of the Roman alphabet mapped to a single key on mobile phone keypads. Most keys have 3 letters associated with them. The user selects among three possible input states corresponding to a closed, slightly open, or open mouth. The letters **s** and **z** occur on 4 letter keys of the keypad. These are selected with a distinctive mouth gesture – in our implementation they were selected by puckering the lips while pressing the appropriate key. With this system, each letter of the alphabet can be entered with a single manual key press. Over the course of our experiment the gains in text entry speed were less sizeable than with Japanese text entry, perhaps because the arbitrary mapping of mouth shape to letter selected requires practice before it is mastered.

## 8  Conclusion

This paper has introduced several projects exploring direct human-computer interaction by using controlled, intentional movements of the face. Results of studies with standard control tasks indicate that simple, rapid, and robust techniques for tracking head movements and mouth shapes afford a high degree of interactive control. We have developed applications to music, digital art creation, and text entry to demonstrate the expressive and communicative propensity of facial actions. We hope this paper will encourage others working in the domain of facial action or expression recognition to consider using their methods to create facial gesture interfaces.

## Acknowledgements

We thank Michael Haehnel, Gamhewage C. De Silva and Chi-Ho Chan for contributions to some of the projects reviewed here. This work was supported in part by the National Institute of Information and Communications Technology.